\documentclass[12pt]{article}
\usepackage{amssymb}
\begin{document}

\begin{center}
{\hbox to\hsize{\hfill September 2007 }}


\vspace*{1 cm}

{\Large \bf

Aharonov-Bohm interactions of a vector unparticle \\}

\bigskip

\bigskip

\bigskip

{\bf Archil Kobakhidze \\}

\smallskip

{ \small \it
School of Physics, Research Centre for High Energy Physics \\ The University of Melbourne, Victoria 3010, Australia \\
E-mail: archilk@physics.unimelb.edu.au
\\}

\vspace*{1.0cm}

{\bf Abstract}\\
\end{center}
{\small \noindent
Recently Georgi argued that  hypothetical conformally invariant 
hidden sector weakly interacting with ordinary particles will have unusual manifestations at low energies in terms 
of effective degrees of freedom called unparticles. In this note we consider Aharonov-Bohm type of interactions  
due to the vector unparticle coupled to elementary fermions. We have found that quantum mechanical phase shift is path-dependent.}


\vspace*{1.0cm}

\baselineskip=16pt

In recent papers \cite{Georgi:2007ek} and \cite{Georgi:2007si}, Georgi argued that a hypothetical hidden sector with an infrared conformal fixed point (see e.g. \cite{Banks:1981nn} as an example of such a theory) can be described by low-energy effective degrees of freedom that have properties qualitatively very different from those of ordinary particles. He named these degrees of freedom unparticles. Among the distinct properties of unparticles are the unusual scaling of the apparent phase space volume, unconventional missing energy spectra and novel interference patterns. These and related issues has been subsequently considered in much details by many authors. All the previous studies have been concentrated on the effects of local interactions of unparticles with the ordinary particles.  In this brief note we would like to discuss Aharonov-Bohm type of interactions \cite{Aharonov:1959fk} associated with a vector unparticle coupled to elementary fermions. 

Following \cite{Georgi:2007ek},\cite{Georgi:2007si}, we consider an effective low-energy interaction of a vector unparticle field ${\cal U}_{\mu}$ (we assume it is transverse, $\partial^{\mu}{\cal U}_{\mu}=0$) with the standard model fermions $f$, 
\begin{equation}
{\cal L}_{\rm int}=\frac{c_f}{\Lambda^{d_{{\cal U}}-1}}\bar \Psi_f\gamma^{\mu}\Psi_f {\cal U}_{\mu}~.
\label{1}
\end{equation}
Here $1<d_{{\cal U}}<2$ is the scaling dimension of the unparticle operator and $\Lambda$ is a mass parameter related to the dynamics of the conformal sector. Unparticle propagator is determined by the scaling dimension \cite{Georgi:2007si},\cite{Cheung:2007ue}, 
\begin{equation}
\Pi^{\mu\nu}(q) = \frac{iA_{d_{{\cal U}}}}{2\sin(\pi d_{{\cal U}})}\left(-\eta^{\mu\nu} + \frac{q^{\mu}q^{\mu}}{q^2}\right)\left(-q^2 \right)^{(d_{{\cal U}}-2)}~,
\label{2}
\end{equation}
where   $A_{d_{{\cal U}}}=\frac{16\pi^{5/2}}{(2\pi)^{2d_{{\cal U}}}}\frac{\Gamma(d_{{\cal U}}+1/2)}{\Gamma(d_{{\cal U}}-1)\Gamma(2d_{{\cal U}})}$.  
In the limit $d_{{\cal U}}\to 1$ one recovers from (\ref{2}) the propagator for the massless vector boson which is  minimally coupled to matter fermions through (\ref{1}). This suggests that unparticle potential will contribute to the standard Aharonov-Bohm phase $\delta_{{\rm AB}}$ \cite{Aharonov:1959fk}, providing it is coupled to electrons, ($c_e\neq 0$ in (\ref{1})). We would like to calculate this contribution. 

Consider the standard Aharonov-Bohm set-up: a long solenoid of radius $R$ carrying a current density $\vec{j}=\vec{\vartheta} B\delta(\rho-R)$ [we use cylindrical coordinates $(\rho, \vartheta, z)$] is placed in the origin of $xy$-plane along the $z$-direction. A coherent beam of charged (with charge $e_f$) particles is split into two parts,  each
moving along $x$-direction of $xy$-plane. The separated beams produce particle
interference pattern on a screen located at some distance from the double slit. Although particles are moving in a force-free region the interference pattern is affected by the magnetic field of the solenoid. One finds the relative phase shift, 
\begin{equation}
\delta_{{\rm AB}}=e_{f}\oint_C\vec{A}\cdot d\vec{x}=e_{f}\Phi
\label{3}
\end{equation}
that depends solely on the magnetic flux $\Phi =\pi R^2B$ ($\vec{A}$ in the above equation is the electromagnetic vector potential, $\vec{B}=\vec{\triangledown}\times \vec{A}$) and not on the path $C$. In order to calculate similar phase shift due to the unparticle vector field, we need to know unparticle vector potential $\vec{{\cal U}}$ produced by the solenoid. Since the problem is stationary we use the propagator (\ref{2}) in the static limit.\footnote{Other stationary potentials due to the exchange of different spin unparticles has been considered in \cite{Liao:2007ic}.} After some calculations we obtain:\footnote{This formulae is valid for $0<d_{{\cal U}}<2$} 
\begin{equation}
U_{\vartheta}(\rho)=\frac{2\pi^2}{(2\pi)^{2d_{{\cal U}}}}\frac{c_eBR}{e\Lambda^{d_{{\cal U}}-1}}\frac{\rho_{<}}{\rho_{>}^{2d_{{\cal U}}-1}}~_2F_1\left((d_{{\cal U}}-1); d_{{\cal U}}; 2; 
\frac{\rho_{<}^2}{\rho_{>}^2}\right)~,
\label{4}
\end{equation}
where "less than" ("greater than") symbol is defined as $\rho_{< (>)}\equiv{\rm min}({\rm max})\left\{\rho, R \right\}$, and $_2F_1$ is the hypergeometric function, $_2F_1(a;b;c;z)=1+\sum_{n=1}^{\infty}\frac{(a)_n(b)_n}{(c)_n n!}$ [$(x)_n=x(x+1)(x+2)...(x+n-1)$ is the rising factorial]. One can easily check that for $d_{{\cal U}}=1$ eq. (\ref{4}) gives the standard result. Since we are interested in the vector potential far outside the solenoid, we take $\rho >> R$ limit in (\ref{4}) and obtain:
\begin{equation}
U_{\vartheta}(\rho)=\frac{2\pi^2}{(2\pi)^{2d_{{\cal U}}}}~\frac{c_eBR^2}{e\Lambda^{d_{{\cal U}}-1}}~\frac{1}{\rho^{2d_{{\cal U}}-1}}~.
\label{5}
\end{equation}     

Now, a particle $f$ moving in the static potential (\ref{5}) is described (in the non-relativistic limit) by the classical Lagrangian:
\begin{equation}
{\cal L}_f=\frac{\vec{p}^2}{2m_f}+\frac{c_f}{m_f\Lambda^{d_{{\cal U}}-1}}\vec{{\cal U}}\cdot\vec{p}~.
\label{6}
\end{equation}
In sharp contrast with the case of an ordinary electromagnetic vector potential, the region outside of soleboid is not force-free. Namely we obtain from (\ref{6}) that the classical motion of a particle outside the solenoid is accelerating, 
\begin{equation}
m_f\dot{\vec{v}}=\frac{c_f}{\Lambda^{d_{{\cal U}}-1}}\vec{v}\times \vec{{\cal B}}~,
\label{a}
\end{equation}
where $\vec{{\cal B}}\equiv \vec{\triangledown}\times \vec{{\cal U}}=\left(0,0, (1-d_{{\cal U}})(2\pi\rho)^{2-2d_{{\cal U}}} \frac{c_eBR^2}{e\Lambda^{d_{{\cal U}}-1}} \right)$. However, the force acting on a particle is centripetal and, as a result, there is no contribution to the Aharonov-Bohm phase associated with the kinetic term in (\ref{6}). Thus, similar to the standard case, the only contribution to the Aharonov-Bohm phase comes from the second term in (\ref{6}). Curiously, this contribution is path-dependent\footnote{Recall, in case of a standard vector potential, the path-dependent phase shift can only emerge when classical force is acting on a particle, e.g. in Lorentz-violating electrodynamics \cite{Kobakhidze:2007iz}, through the change in particle momentum.}:
\begin{equation}
\delta_{{\cal U}}=\frac{2\pi^2}{(2\pi)^{2d_{{\cal U}}}}~\frac{c_fc_e}{e}~\frac{BR^2}{\Lambda^{2d_{{\cal U}}-2}}
\oint_{C} \rho^{2-2d_{{\cal U}}}d\vartheta
\label{7}
\end{equation}  
In the standard case ($d_{{\cal U}}=1$) (\ref{7}) reduces to the path-independent topological phase similar to the one in eq. (\ref{3}). 

To estimate Aharonov-Bohm phase due to the unparticle vector potential we consider two representative paths. First, suppose $C$ is a circle of a radius $r$, $\rho=r=$const. Upon the integration in (\ref{7}) we obtain:
\begin{equation}
\delta_{{\cal U}}=\frac{1}{(2\pi)^{2d_{{\cal U}}-2}}~\frac{c_fc_e}{e}~\frac{1}{\left( r\Lambda \right)^{2d_{{\cal U}}-2}}~\Phi
\label{8}
\end{equation}
As a second example, consider a classical path with the length along $x$-direction much longer than the length in $y$-direction. Then we approximate, $\rho_{1,2}=\frac{a_{1,2}}{\sin\vartheta}$, where $a_1$ and $a_2$ are the distances from the solenoid to the first and second beams, 
respectively. The phase (\ref{7}) takes the form:
\begin{equation}
\delta_{{\cal U}}=\frac{\sqrt{\pi}\Gamma(d_{{\cal U}}-1/2)}{(2\pi)^{2d_{{\cal U}}-1}\Gamma(d_{{\cal U}})}~\frac{c_fc_e}{e}~
\left( \frac{1}{\left( a_1\Lambda \right)^{2d_{{\cal U}}-2}}+\frac{1}{\left( a_2\Lambda \right)^{2d_{{\cal U}}-2}}\right)~\Phi
\label{9}
\end{equation} 

To conclude, we have found that Aharonov-Bohm phase shift induced by a hypothetical unparticle vector potential  is path-dependent.  This striking signature potentially can be tested in tabletop interference experiments. If $f$ is electrically charged fermion (e.g. electron) then the path-dependent phase will be superimposed on the standard path-independent Aharonov-Bohm phase. However, in the case of one solenoid considered in this paper, the extra path-dependent phase shift is expected to be much smaller than the standard topological phase. Indeed, if we take, e.g. $\Lambda \approx 1$TeV, $c_e\approx e$ and $r\approx 0.01$cm in (\ref{8}), we estimate $\delta_{{\cal U}}\approx 10^{-36(d_{\cal U}-1)}\delta_{{\rm AB}}$. With currently available experimental techniques such small phase shift 
can be detected only if $d_{{\cal U}}\lesssim 1.06$. However, the effect of unparticle can be significantly enhanced, for example, by considering an array of solenoids. On the other hand, if the fermion $f$ is a neutral particle (e.g. neutron), only path-dependent phase shift is expected. Moreover, instead of magnetic solenoid one can use a rotating rod made of material with excess of neutrons (e.g. the quartz rod). We hope to return to more detailed analysis of such experiments in future.

\subparagraph{Acknowledgment.}

This work was supported by the Australian Research Council.

\end{document}